\documentclass[10pt,journal,compsoc]{IEEEtran}
\usepackage[pdftex]{graphicx,xcolor}
\usepackage{url}

\usepackage[fleqn]{amsmath} 
\usepackage{newtxtext} 
\usepackage[varg]{newtxmath} 

\usepackage{latexsym}
\usepackage{verbatim}
\usepackage{subfigure}
\usepackage[all]{xy}
\usepackage[none]{hyphenat}

\usepackage{bbding}
\usepackage{pifont}
\usepackage{wasysym}
\usepackage{amssymb}

\newif\iftodo
\todotrue

\newif\iftable
\tabletrue

\newif\ifmemo
\memofalse

\newif\ifmark
\markfalse

\newif\ifrelatedwork
\relatedworkfalse

\newtheorem{definition}{Definition}

\newcommand\tablength{0.3}
\newcommand\figwidth{80mm}

\newcommand\htab{\hspace{\tablength cm}}

\ifCLASSOPTIONcompsoc
  \usepackage[nocompress]{cite}
\else
  \usepackage{cite}
\fi
\ifCLASSINFOpdf
\else
\fi
\begin{document}
%
\title{A Deep Reinforcement Learning Based Approach\\
 for Cost- and Energy-Aware Multi-Flow Mobile Data Offloading}
%
%
%
%
\author{Cheng~Zhang$^1$,~
        Zhi~Liu$^2$,~
        Bo~Gu$^3$,~
        Kyoko~YAMORI$^4$,~
        and~Yoshiaki~TANAKA$^5$\\
        \footnotesize{$^1$Department of Computer Science and Communications Engineering,
Waseda University, Tokyo, 169-0072 Japan,Email: cheng.zhang@akane.waseda.jp\\
       $^2$Department of Mathematical and Systems Engineering, Shizuoka University, Shizuoka, 432-8561 Japan\\
       $^3$Department of Information and Communications Engineering,
Kogakuin University, Tokyo, 192-0015 Japan\\
        $^4$Department of Management Information, Asahi University, Mizuho-shi, 501-0296 Japan\\
        $^5$Department of Communications and Computer Engineering,
Waseda University, Tokyo, 169-8555 Japan}

}

\IEEEtitleabstractindextext{%
\begin{abstract}
With the rapid increase in demand for mobile data, mobile network 
operators are trying to expand wireless network capacity by deploying 
wireless local area network (LAN) hotspots on to which they can offload 
their mobile traffic. However, these network-centric methods usually 
do not fulfill the interests of mobile users (MUs). Taking into consideration 
many issues such as different applications' deadlines, monetary cost 
and energy consumption, how the MU decides whether to offload their 
traffic to a complementary wireless LAN is an important issue. 
Previous studies assume the MU's mobility pattern is known in advance, 
which is not always true.  In this paper, we study the MU's policy to
minimize his monetary cost and energy consumption without known MU 
mobility pattern.  We propose to use a kind of reinforcement learning 
technique called deep Q-network (DQN) for MU to learn the optimal 
offloading policy from past experiences. In the proposed DQN based 
offloading algorithm, MU's mobility pattern is no longer needed. 
Furthermore, MU's state of remaining data is directly fed into the 
convolution neural network in DQN without discretization. Therefore, 
not only does the discretization error present in previous work disappear, but also 
it makes the proposed algorithm has the ability to generalize the past 
experiences, which is especially effective when the number of 
states is large. Extensive simulations are conducted to validate 
our proposed offloading algorithms.
\end{abstract}

}

\maketitle

\IEEEdisplaynontitleabstractindextext

%
\IEEEpeerreviewmaketitle

\IEEEraisesectionheading{\section{Introduction}\label{sec:introduction}}
  \IEEEPARstart{T}{he} mobile data traffic demand is growing rapidly. According to the
  investigation of Cisco Systems \cite{CiscoVNI2015}, the mobile data
  traffic is expected to reach 24.3 exabytes per month by 2019, while
  it was only 2.5 exabytes per month at the end of 2014.  On the other
  hand, the growth rate of the mobile network capacity is far from
  satisfying that kind of the demand, which has become a major problem
  for wireless mobile network operators (MNOs). Even though 5G
  technology is promising for providing huge wireless network
  capacity \cite{5GCapacity2014}, the development process is long
  and the cost is high. Economic methods such as time-dependent
  pricing \cite{TDPDuopolyNSP}\cite{TDPOligopolyNSP}
  have been proposed to change users' usage pattern, which are
  not user-friendly. Up to now, the best practice for increasing mobile
  network capacity is to deploy complementary networks (such as
  wireless LAN and femtocells), which can be quickly deployed and
  is cost-efficient. Using such methods, part of the MUs' traffic
  demand can be offloaded from a MNO's cellular network to the
  complementary network.\\
  \indent The process that a mobile device automatically changes
  its connection type (such as from cellular network to wireless LAN)
  is called \emph{vertical handover} \cite{Marquez-Barja:2011}.
  Mobile data offloading is facilitated by new standards such as
  Hotspot 2.0 \cite{Hotspot20} and the 3GPP Access Network Discovery
  and Selection Function (ANDSF) standard \cite{3GPP4WiFi}, with
  which information of network (such as price and network load) can
  be broadcasted to MUs in real-time. Then MUs can make offloading
  decisions intelligently based on the real-time network information.\\
   \indent There are many works related to the wireless LAN offloading
   problem. However, previous works either considered the wireless
   LAN offloading problem from the network providers' perspective
   without considering the MU's quality of service (QoS)
   \cite{BargainingOffloading}\cite{DoubleAuction2014}, or studied
   wireless LAN offloading from the MU's perspective
   \cite{Aug3GWiFi2010}\cite{LeeHowMuchWiFi2013}
   \cite{AMUSEMungChiang2013}\cite{DAWNHuang2015},
   but without taking the energy consumption as well as cost problems
   into consideration.\\
  \indent \cite{ChengAPNOMS2016}\cite{MDPOffloading2017}
  studied wireless LAN offloading problem from MU's perspective.
  While single-flow mobile date offloading was considered
  in \cite{ChengAPNOMS2016}, multi-flow mobile data offloading 
  problem is studied in \cite{MDPOffloading2017} in which a MU 
  has multiple applications to transmit data simultaneously with 
  different deadlines. MU's target was to minimize its total cost, 
  while taking monetary cost, preference for energy consumption, 
  application's delay tolerance into
  consideration. This was formulated the wireless LAN offloading 
  problem as
  a finite-horizon discrete-time Markov decision process
  \cite{LiuZhiAccess}\cite{LiuZhiConfMDP2011}\cite{LiuZhiTrans2013}.
  A high time complexity dynamic programming (DP) based optimal
  offloading algorithm and a low time complexity heuristic offloading
  algorithm were prosed in \cite{MDPOffloading2017}.\\
\begin{table*}[t]
\caption{Comparison of different works.}
\centering
\label{tablerelatedwork}
\renewcommand{\arraystretch}{1}
\begin{tabular}{|c|c|c|c|c|c|}
\hline

\hline
References & Multi-flow & Unknown mobility pattern & No discretization error & Energy consideration & Q-value prediction\\
\hline
\cite{DAWNHuang2015} & $\times$  
& $\times$  & $\times$ & $\times$ & --- \\
\hline
\cite{ChengAPNOMS2016} & $\times$  
& $\checkmark$  & $\times$ & $\checkmark$ & $\times$ \\
\hline
\cite{MDPOffloading2017} & $\checkmark$  
& $\times$  & $\times$ & $\checkmark$ & ---\\
\hline
This paper & $\checkmark$  
& $\checkmark$  & $\checkmark$ & $\checkmark$ & $\checkmark$\\

\hline

\hline
\end{tabular}
\end{table*}
  \indent One assumption in \cite{MDPOffloading2017}
  was that MU's mobile pattern from one place to another is 
  known in advance, then the transition probability, which 
  is necessary for optimal policy calculation in MDP, can be 
  obtained in advance.  However, the MU's mobility pattern 
  may not be easily obtained or the accuracy is not high.
  Even though a Q-learning \cite{RLBook} based algorithm was proposed
  in \cite{ChengAPNOMS2016} for the unknown MU's mobility pattern case,
  the learning, the convergence rate of the proposed algorithm is rather low
  due to the large number of states. It takes time for the reinforcement learning
  agent to experience all the states to estimate the Q-value.\\
  \indent In this paper, we propose a deep reinforcement learning algorithm,
  specifically, a deep Q-network (DQN) \cite{DeepRL} based algorithm, 
  to solve the multi-flow offloading problem with high convergence rate
  without knowing MU's mobility pattern.  In reinforcement learning,
  the agent learns to make optimal
decisions from past \textit{experience}
   when interacting with the \textit{environment} (see Fig. 1).
  In the beginning, the agent has no knowledge of the task and 
  makes decision (or takes \textit{action}),
  then, it receives a \textit{reward} based on how well the task  is done. 
  Theoretically, if the agent experience all the situations
  (\textit{states}) and get to know the value of its decision on all 
  situations, the agent can make optimal decisions. However,
 often it is impossible for the agent to experience all situations.
  The agent does not have the ability to generalize its
  experience when unknown situations appear. 
  Therefore, DQN \cite{DeepRL} was developed to let agent 
  generalize its experience. DQN uses deep neural networks 
  (DNN) \cite{Deeplearning} to predict the Q-value in standard
  Q-learning, which is a value that maps from agent's state 
  to different actions. Then, the optimal offload action can 
  be directly obtained by choosing the one that has the 
  maximum Q-value. Please refer to Table. \ref{tablerelatedwork}
  for the comparison of different works.\\
  \indent The rest of this paper is organized as follows.
  Section \ref{systemmodel} illustrates the system model.
  Section \ref{probform} defines MU's mobile data offloading
  optimization target.  Section \ref{learningalgorithm}
  proposes DQN based algorithm.
  Section \ref{performanceevaluation} illustrates the simulation
  and results.
  Finally, we conclude this paper in Section \ref{conclusion}.

\section{System Model}\label{systemmodel}
  Since the cellular network coverage is rather high,
  we assume the MU can always access the cellular network, 
  but cannot always access wireless LAN. The wireless LAN 
  access points (APs) are usually deployed at home, stations, 
  shopping malls and so on. Therefore, we assume that wireless 
  LAN access is location-dependent. We mainly focus on applications 
  with data of relative large size and delay-tolerance to download, 
  for example,  applications like software updates, file downloads.
  The MU has $M$ files to download from a remote server.
  Each file forms a flow, and the set of flows is denoted as
  $\mathcal{M}$$=$$\{1,...,M\}$. Each flow $j\in\mathcal{M}$ has
  a deadline $T^j$. $\textbf{\textit{T}}$$=$$(T^1, T^2, ... , T^M)$ is
  the deadline vector for the MU's $M$ flows. Without loss of generality,
  it is assumed that $T^1 \le T^2 \le  ... \le T^M$. We consider a slotted
  time system as $t$$\in$$\mathcal{T}$$=$$\{1,...,T^M\}$.
  
  \ifmark
  \textcolor{red}{We only considered delay-tolerant traffic in this paper
  in which $T^j > 0$ for $j\in\mathcal{M}$. For non-delay-tolerant 
  traffic, the deadline $T^j = 0$. In this case, MU has to start to 
  transmit data whenever there is a network available without network
  selection.}
  \else
  We only considered delay-tolerant traffic in this paper
  in which $T^j > 0$ for $j\in\mathcal{M}$. For non-delay-tolerant 
  traffic, the deadline $T^j = 0$. In this case, MU has to start to 
  transmit data whenever there is a network available without network
  selection.
  \fi
  
  To simplify the analysis, we use limited discrete locations instead
  of infinite continuous locations. It is assumed that a MU can move
  among the $L$ possible locations, which is denoted as
  set $\mathcal{L}$$=$$\{1,...,L\}$.
  While the cellular network is available at all the locations, the
  availability of wireless LAN network is dependent on
  location $l\in\mathcal{L}$. The MU has to make a decision on
  what network to select and how to allocate the available data rate
  among $M$ flows at location $l$ at time $t$ by considering
  total monetary cost, energy consumption and remaining time
  for data transmission.
  \ifmark
  \textcolor{red}{The consideration of MU's energy consumption
  is one feature of our series of works
  \cite{ChengAPNOMS2016}\cite{MDPOffloading2017}.
  When the data rate of wireless LAN is too low, the energy consumption
  per Mega bytes data is high \cite{EnergyThroughput} (see Fig.\ref{energythroughput} ) and it will take a long time to transmit MU's data. 
  For MUs who care about energy consumption or have a short deadline 
  for data transmission, they may choose to use cellular network with 
  high data rate even if wireless LAN is available.}
  \else
   The consideration of MU's energy consumption
  is one feature of our series of works
  \cite{ChengAPNOMS2016}\cite{MDPOffloading2017}.
  When the data rate of wireless LAN is too low, the energy consumption
  per Mega bytes data is high \cite{EnergyThroughput} (see Fig.\ref{energythroughput} ) and it will take a long time to transmit MU's data. 
  For MUs who care about energy consumption or have a short deadline 
  for data transmission, they may choose to use cellular network with 
  high data rate even if wireless LAN is available.
  \fi
  As in \cite{DAWNHuang2015}\cite{ChengAPNOMS2016}\cite{MDPOffloading2017}
  the MU's decision making problem can be modeled as a
  finite-horizon Markov decision process.\\
  \begin{figure}[t]
   \centering
   \includegraphics[width=80mm]{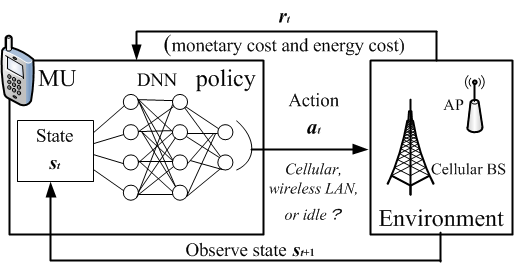}
   \caption{An deep Q-network based modeling: at time decision epoch $t$,
   the \textit{state} of MU contains location $l$ and remaining file size B,
   which is fed into a deep neural network to generate optimal policy.
   Then MU chooses actions of \textit{Wireless LAN}, \textit{Cellular network},
   or \textit{Idle}, which incur different cost on MU. The objective of MU is
   to minimize total cost from time $1$ to $T$.}\label{scenarioFig}
   \end{figure}
  \indent We define the system \textit{state} at $t$ as in Eq. (\ref{state})
  \begin{equation}\label{state}
    \textbf{\textit{s}}_t = \{l_t,\textbf{\textit{b}}_t\}
  \end{equation}
  where $l_t\in \mathcal{L}$$=$$\{1,...,L\}$ is the MU's
  location index at time $t$, which can be obtained from GPS.
  $\mathcal{L}$ is the location set.
  $\textbf{\textit{b}}_t$$=$$(b_t^1, b_t^2, ... , b_t^M)$ is the vector of
  remaining file sizes of all $M$ flows at time $t$, $b_t^j\in \mathcal{B}^j$
  $=[0,B^j]$ for all $j\in \mathcal{M}$.
  \ifmark
  \textcolor{blue}{
    $B^j$ is the total file size to be transmitted on flow $j$.
    $b_t^j$ is equal to $B^j$ before flow $j$ starts to transmit. 
    $\mathcal{B}$=$(\mathcal{B}^1,\mathcal{B}^2, ... , \mathcal{B}^M)$
    is the set of vectors.}\\
  \else
    $B^j$ is the total file size to be transmitted on flow $j$.
    $b_t^j$ is equal to $B^j$ before flow $j$ starts to transmit. 
    $\mathcal{B}$=$(\mathcal{B}^1,\mathcal{B}^2, ... , \mathcal{B}^M)$
    is the set of vectors.\\
  \fi
  \indent The MU's \textit{action} $a_t$ at each decision epoch $t$
  is to determine whether to transmit data through wireless LAN
  (if wireless LAN is available), or cellular network, or just keep idle
  and how to allocate the network data rate to $M$ flows. 
\ifmark
\textcolor{red}{Please note that epoch is the same as time slot, 
at which MU makes action decision. We use epoch and 
time slot interchangeably in this paper. The reason why MU 
does not choose any of the network  is that MU may 
try to wait for free or low price wireless 
LAN to save his/her money even though cellular is ready anytime.
The survey in \cite{Soumya2013} had the results that more than 
50\% of the respondents would like to wait for 10 minutes to 
stream YouTube videos and 3-5 hours to download a file a 
monetary incentive is given.}
 \else
{Please note that epoch is the same as time slot, 
at which MU makes action decision. We use epoch and 
time slot interchangeably in this paper. 
The reason why MU does not choose any of the 
network  is that MU may try to wait for free or low price wireless 
LAN to save his/her money even though cellular is ready anytime.
The survey in \cite{Soumya2013} had the results that more than 
50\% of the respondents would like to wait for 10 minutes to 
stream YouTube videos and 3-5 hours to download a file a 
if monetary incentive is given.}
 \fi  
\ifmark
\textcolor{red}{The reason why MU does not choose any of the 
network  is that MU may try to wait for free or low price wireless 
LAN to save his/her money even though cellular is ready anytime.
The survey in \cite{Soumya2013} had the results that more than 
50\% of the respondents would like to wait for 10 minutes to 
stream YouTube videos and 3-5 hours to download a file a 
monetary incentive is given.}
 \else
{The reason why MU does not choose any of the 
network  is that MU may try to wait for free or low price wireless 
LAN to save his/her money even though cellular is ready anytime.
The survey in \cite{Soumya2013} had the results that more than 
50\% of the respondents would like to wait for 10 minutes to 
stream YouTube videos and 3-5 hours to download a file a 
if monetary incentive is given.}
 \fi

  Therefore, the MU's action vector is denoted as in Eq. (\ref{action})
  \begin{equation}\label{action}
      \textbf{\textit{a}}_{t}=(\textbf{\textit{a}}_{t,c},\textbf{\textit{a}}_{t,w})
  \end{equation}
  where $\textbf{\textit{a}}_{t,c}=(a_{t,c}^1, a_{t,c}^2, ..., a_{t,c}^M)$
  denotes the vector of cellular network allocated data rates,
  $a_{t,c}^j$  denotes the cellular data rate allocated to flow
  $j\in$$\mathcal{M}$, and
  $\textbf{\textit{a}}_{t,w}=(a_{t,w}^1, a_{t,w}^2, ... , a_{t,w}^M)$ denotes
  the vector of wireless LAN network allocated data rates, and $a_{t,w}^j$
  denotes the wireless LAN rate allocated to flow $j\in$$\mathcal{M}$.
  Here the subscript $c$ and $w$ stand for cellular network and
  wireless LAN, respectively. Please note
  that $a_{t,w}^1$, $a_{t,w}^2$, ..., $a_{t,w}^M$
  all can be 0 if the MU is not in the coverage area of wireless LAN AP.
  Even though it is technically possible that wireless LAN and cellular
  network can be used at the same time, we assume that the MU can
  not use wireless LAN and cellular network at the same time.
  We make this assumption for two reasons:
  (i) If we restrict the MU to use only one network interface at the same
  time slot, then the MU's device may be used for longer time with the
  same amount of left battery.
  (ii) Nowadays smartphones, such as an iPhone, can only use one
  network interface at the same time. We can easily implement our algorithms
  on a MU's device without changing the hardware or OS of the smartphone
  if we have this assumption.
  At time $t$, MU may choose to use wireless LAN (if wireless LAN
  is available) or cellular network, or not to use any network. If the MU
  chooses wireless LAN at $t$, the wireless LAN network allocated data
  rate to flow $j$, $a_{t,w}^j$, is greater than or equal to 0, and the MU does not use
  cellular network in this case, then $a_{t,c}^j$ = 0.  On the other hand,
  if the MU chooses cellular network at $t$, the cellular network allocated
  data rate to flow $j$, $a_{t,c}^j$, is greater than or equal to 0, and the MU does not
   use wireless LAN in this case, then $a_{t,w}^j$ = 0.
  $a_{t,n}^j$, $n\in\{c,w\}$ should not be greater than the remaining file
  size $b_t^j$ for flow $j\in$$\mathcal{M}$.\\
  \indent The sum data rate of all the flows of cellular network and
  wireless LAN are denoted as $a_{t,c}=\sum_{j\in \mathcal{M}}a_{t,c}^j$
  and $a_{t,w}=\sum_{j\in\mathcal{M}}a_{t,w}^j$, respectively.
  $a_{t,c}$ and $a_{t,w}$ should satisfy the following conditions.
  \begin{equation}
  a_{t,c} \le \gamma_{c}^l
  \end{equation}
  \begin{equation}
  a_{t,w} \le \gamma_{w}^l
  \end{equation}
  $\gamma_{c}^l$ and $\gamma_{w}^l$ are the maximum data rates of cellular network and
  wireless LAN, respectively, at each location $l$.\\
\iftable
\begin{table}[t]\label{table1}
\caption{Notations summary.}
\renewcommand{\arraystretch}{1}
\begin{tabular}{cp{6.3cm}}
\hline

\hline
Notation & Description\\
\hline

\hline
$\mathcal{M}$ & $\mathcal{M}$$=$$\{1,...,M\}$, MU's $M$ flows set. \\
$\textbf{\textit{T}}$ & $\textbf{\textit{T}}$$=$$(T^1, T^2, ... , T^M)$, MU's deadline vector. \\
$t$ & $t\in\mathcal{T}^M$, the specific decision epoch of MU. \\
$\mathcal{L}$ & $\mathcal{L}$$=$$\{1,...,L\}$, the location set of MU. \\
$\mathcal{B}^j$ & $\mathcal{B}^j$$\subseteq$$[0,...,b_t^j]$, the total size of MU's $j$ flow. $j\in \mathcal{M}$. \\
$\textbf{\textit{b}}_t$ & $\textbf{\textit{b}}_t$$=$$(b_t^1, b_t^2, ... , b_t^M)$, vector of remaining file size. \\
$\textbf{\textit{s}}_t$ & $\textbf{\textit{s}}_t=(l_t,\textbf{\textit{b}}_t)$, state of MU.\\
$l_t$   & $l_t\in \mathcal{L}$, MU's location index at time $t$.\\
$a_{t,c}^j$  & cellular data rate allocated to flow $j\in$$\mathcal{M}$ at time $t$\\
$a_{t,w}^j$ &  wireless LAN data rate allocated to flow $j\in$$\mathcal{M}$ at time $t$\\
$\textbf{\textit{a}}_{t,c}$&$\textbf{\textit{a}}_{t,c}=\{a_{t,c}^1, a_{t,c}^2, ..., a_{t,c}^M\}$\\
$\textbf{\textit{a}}_{t,w}$&$\textbf{\textit{a}}_{t,w}=\{a_{t,w}^1, a_{t,w}^2, ..., a_{t,w}^M\}$\\
$\textbf{\textit{a}}_{t}$ &$\textbf{\textit{a}}_{t}=(\textbf{\textit{a}}_{t,c},\textbf{\textit{a}}_{t,w})$\\
$ a_{t,c}$&$ a_{t,c}= \sum_{j\in \mathcal{M}} a_{t,c}^j$\\
$a_{t,w}$&$ a_{t,w}= \sum_{j\in \mathcal{M}} a_{t,w}^j$\\
$\gamma_{c}^l$ & cellular throughput in bps at location $l$.\\
$\gamma_{w}^l$ & wireless LAN throughput in bps at location $l$.\\
$\varepsilon_c^l$ & energy consumption rate of celllar network  in joule/bits at location $l$.\\
$\varepsilon_w^l$ & energy consumption rate of wireless LAN in joule/bits at location $l$.\\
$\theta_t$ & energy preference of MU at $t$.\\
$p_c$ & MNO's usage-based price for cellular network service.\\
$\hat{c}_{T^M+1}(\cdot)$&MU's penalty function for remaining data at $T^M+1$.\\
$\xi_t(\textbf{\textit{s}}_t,\textbf{\textit{a}}_t)$& MU's energy consumption at $t$.\\
$\phi_t$ & $\mathcal{L}$$\times$$\mathcal{K}$$\rightarrow$$\mathcal{A}$, transmission decision at $t$.\\
$\pi$ & $\pi=\{\phi_t(l,b),\;\forall\; t\in\mathcal{T^M}, l\in\mathcal{L}, b\in\mathcal{B}\}$, MU's policy.\\
\hline

\hline
\end{tabular} \\
\end{table}
\else
\fi
  \indent At each epoch $t$, three factors affect the MU's decision.
  \begin{description}

    \item (1) \textit{monetary cost}: it is the payment from the MU
    to the network service provider. We assume that the network
    service provider adopts $\textit{usage-based pricing}$, which
    is being widely used by carriers in Japan, USA, etc.
    The MNO's price is denoted as $p_c$. It is assumed that
    wireless LAN is free of charge. We define the monetary cost
    $c_t(\textbf{\textit{s}}_t,\textbf{\textit{a}}_t)$ as in Eq. (\ref{payment})
    \ifmark
    \textcolor{blue}{
    \begin{equation}\label{payment}
       c_t(\textbf{\textit{s}}_t,\textbf{\textit{a}}_t)= p_c\sum_{j\in\mathcal{M}}\min\{b_t^j,\delta a_{t,c}^j\}
    \end{equation}
    }
    \else
    \begin{equation}\label{payment}
       c_t(\textbf{\textit{s}}_t,\textbf{\textit{a}}_t)= p_c\sum_{j\in\mathcal{M}}\min\{b_t^j,\delta a_{t,c}^j\}
    \end{equation}
    \fi
    \item (2) \textit{energy consumption}: it is the energy consumed
    when transmitting data through wireless LAN or cellular network.
    We denote the MU's awareness of energy as in Eq. (\ref{energy})
    \ifmark
    \textcolor{blue}{
    \begin{equation}\label{energy}
    \begin{split}
        \xi_t(\textbf{\textit{s}}_t,\textbf{\textit{a}}_{t})
        =&\theta_t (\varepsilon_c^l\sum_{j\in\mathcal{M}}\min\{b_t^j,\delta a_{t,c}^j\}\\
        &+ \varepsilon_w^l\sum_{j\in\mathcal{M}}\min\{b_t^j,\delta a_{t,w}^j\})
    \end{split}
    \end{equation}
    }
    \else
    \begin{equation}\label{energy}
    \begin{split}
        \xi_t(\textbf{\textit{s}}_t,\textbf{\textit{a}}_{t})
        =&\theta_t (\varepsilon_c^l\sum_{j\in\mathcal{M}}\min\{b_t^j,\delta a_{t,c}^j\}\\
        &+ \varepsilon_w^l\sum_{j\in\mathcal{M}}\min\{b_t^j,\delta a_{t,w}^j\})
    \end{split}
    \end{equation}
    \fi
    
    where $\varepsilon_c^l$ is the energy consumption rate of the
    cellular network in joule/bits at location $l$ and $\varepsilon_w^l$
    is the energy consumption rate of the wireless LAN in joule/bits
    at location $l$. It has been shown in \cite{EnergyCollaborate2015}
    that both $\varepsilon_c^l$ and $\varepsilon_w^l$ decrease with
    throughput, which means that low transmission speed consumes
    more energy when transmitting the same amount of data.
    According to \cite{EnergyDownUplink}, the energy consumptions
    for downlink and uplink are different. Therefore, the energy
    consumption parameters $\varepsilon_c^l$ and  $\varepsilon_w^l$
    should be differentiated for downlink or uplink, respectively.
    In this paper, we do not differentiate the parameters for downlink or
    uplink because only the downlink case is considered.  Nevertheless,
    our proposed algorithms are also applicable for uplink scenarios with
    energy consumption parameters for uplink.
    $\theta_t$ is the MU's preference for energy consumption at time $t$.
    $\theta_t$ is the weight on energy consumption set by the MU. Small
    $\theta_t$ means that the MU cares less on energy consumption. For
    example, if the MU can soon charge his smartphone, he may set $\theta_t$
    to a small value, or if the MU is in an urgent status and could not
    charge within a short time, he may set a large value for $\theta_t$.
    $\theta_t$ = 0 means that the MU does not consider energy consumption
    during the  data offloading
    \item (3) \textit{penalty}: if the data transmission does not finish
    before deadline $T^j$, $j\in\mathcal{M}$, the penalty for the MU is
    defined as Eq. (\ref{penalty}).
    \begin{equation}\label{penalty}
        \hat{c}_{T^j+1}(\textbf{\textit{s}}_{T^j+1})=\hat{c}_{T^j+1}(l_{T^j+1},\textbf{\textit{b}}_{T^j+1})=g(\textbf{\textit{b}}_{T^j+1})
    \end{equation}
    where $g(\cdot)$ is a non-negative non-decreasing function.
    ${T^j+1}$ means that the penalty is calculated after deadline
    $T^j$.
  \end{description}

 \section{Problem Formulation}\label{probform}
  \ifmark
  \textcolor{blue}{MU has to decide all the actions from the first 
  time epoch to the last one to minimize overall monetary cost 
  and energy consumption for all the time epochs. Policy is 
  the sequence of actions from the first time epoch to the last one. 
  We formally have definition for \textit{policy} as follows.
  \begin{definition}
  The MU's \textit{policy} is the actions he takes
  from $t=1$ to $t=T^M$, which is defined as in Eq. (\ref{policy})
 \begin{equation}\label{policy}
   \pi=\bigg\{\phi_t(l_t,\textbf{\textit{b}}_t),\;\forall\; t\in\mathcal{T},l\in\mathcal{L}, \textbf{\textit{b}}_t\in\mathcal{B}\bigg\}
 \end{equation}
 where $\phi_t(l_t,\textbf{\textit{b}}_t)$ is a function mapping from state
 $\textbf{\textit{s}}_t=(l_t,\textbf{\textit{b}}_t)$ to a decision action at $t$.
  \end{definition}
  }
  \else
  MU has to decide all the actions from the first 
  time epoch to the last one to minimize overall monetary cost 
  and energy consumption for all the time epochs. Policy is 
  the sequence of actions from the first time epoch to the last one. 
  We formally have definition for \textit{policy} as follows.
  \begin{definition}
  The MU's \textit{policy} is the actions he takes from
  from $t=1$ to $t=T^M$, which is defined as in Eq. (\ref{policy})
 \begin{equation}\label{policy}
   \pi=\bigg\{\phi_t(l_t,\textbf{\textit{b}}_t),\;\forall\; t\in\mathcal{T},l\in\mathcal{L}, \textbf{\textit{b}}_t\in\mathcal{B}\bigg\}
 \end{equation}
 where $\phi_t(l_t,\textbf{\textit{b}}_t)$ is a function mapping from state
 $\textbf{\textit{s}}_t=(l_t,\textbf{\textit{b}}_t)$ to a decision action at $t$.
  \end{definition}
  \fi
 The set of $\pi$ is denoted as $\Pi$. If policy $\pi$
 is adopted, the state is denoted as $\textbf{\textit{s}}_t^{\pi}$.

 The objective of the MU is to the minimize the expected
 total cost (include the monetary cost and the energy consumption)
 from $t=1$ to $t=T^M$ and penalty at $t=$${T^M+1}$ with an
 a optimal $\pi^*$ (see Eq. (\ref{totalcost}))
 \begin{equation}\label{totalcost}
    \min_{\pi\in\Pi} E_{\textbf{\textit{s}}_1}^{\pi}
    \Bigg[
    \sum_{t=1}^{T^M}r_t(\textbf{\textit{s}}_t^{\pi},\textbf{\textit{a}}_{t})
    +\sum_{j\in\mathcal{M}}\hat{c}_{T^j+1}(\textbf{\textit{s}}_{T^j+1}^{\pi})
    \Bigg]
\end{equation}
where $r_t(\textbf{\textit{s}}_t,\textbf{\textit{a}}_{t})$ is the sum of the
 monetary cost and the energy consumption as in Eq. (\ref{reward})
 \begin{equation}\label{reward}
    r_t(\textbf{\textit{s}}_t,\textbf{\textit{a}}_{t})=c_t(\textbf{\textit{s}}_t,\textbf{\textit{a}}_{t})+\xi_t(\textbf{\textit{s}}_t,\textbf{\textit{a}}_{t})
 \end{equation}
\ifmark
\indent  \textcolor{red}{The \textit{optimal policy} is the the optimal solution
  of the minimization problem defined in Eq. (\ref{totalcost}).}
  \else
\indent  The \textit{optimal policy} is the the optimal solution
  of the minimization problem defined in Eq. (\ref{totalcost}).
  \fi 
Please note that the optimal action at each $t$ does not lead
to the optimal solution for the problem in Eq. (\ref{totalcost}).  At each
time $t$, not only the cost for the current time $t$ should be considered,
but also the future expected cost. \\
\ifmark
\indent  \textcolor{red}{The objective function of the minimization 
problem Eq. (\ref{totalcost}) includes three parts. The first two parts are
denoted in $r_t(\textbf{\textit{s}}_t,\textbf{\textit{a}}_{t})$, which contains
monetary cost and energy consumption. As shown in Eq. (\ref{payment}),
the monetary cost is determined by the cellular network price $p_c$ and 
the data transmitted through cellular network $a_{t,c}^j$. The parameter
$a_{t,c}^j$ is the one that MU tries to determine in each time epoch $t$,
which has the priority all the time. On the other hand, as shown in the 
Eq. (\ref{energy}), the parameters to minimize for energy consumption
are $a_{t,c}^j$ and $a_{t,w}^j$, which are the data transmitted through
cellular network and wireless LAN, respectively. Whether the energy 
consumption has priority is determined by the MU's preference 
parameter $\theta_t$. Obviously, if $\theta_t$ is set to zero by MU, 
Eq. (\ref{energy}) become zero and there is no priority for energy 
consumption in the target minimization problem. In this case,
the value the minimization in Eq. (\ref{totalcost}) reached is the 
minimized  total monetary cost.  If $\theta_t$ is set 
to nonzero by MU,  Eq. (\ref{energy}) is also nonzero and energy 
consumption is also incorporated in the target minimization problem.
In this case, the value the minimization in Eq. (\ref{totalcost}) reached 
is the minimized  total monetary cost and energy consumption.
The third part is the penalty defined in Eq. (\ref{penalty}). The remaining 
data after deadline determines the penalty, which can not be directly 
controlled by MU. MU can only try to finish the data transmission
before the deadline to eliminate the penalty.}
\else
\indent  The objective function of the minimization 
problem Eq.(\ref{totalcost}) includes three parts. The first two parts are
denoted in $r_t(\textbf{\textit{s}}_t,\textbf{\textit{a}}_{t})$, which contains
monetary cost and energy consumption. As shown in Eq. (\ref{payment}),
the monetary cost is determined by the cellular network price $p_c$ and 
the data transmitted through cellular network $a_{t,c}^j$. The parameter
$a_{t,c}^j$ is the one that MU tries to determine in each time epoch $t$,
which has the priority all the time. On the other hand, as shown in the 
Eq. (\ref{energy}), the parameters to minimize for energy consumption
are $a_{t,c}^j$ and $a_{t,w}^j$, which are the data transmitted through
cellular network and wireless LAN, respectively. Whether the energy 
consumption has priority is determined by the MU's preference 
parameter $\theta_t$. Obviously, if $\theta_t$ is set to zero by MU, 
Eq. (\ref{energy}) become zero and there is no priority for energy 
consumption in the target minimization problem. In this case,
the value the minimization in Eq. (\ref{totalcost}) reached is the 
minimized  total monetary cost.  If $\theta_t$ is set 
to nonzero by MU,  Eq. (\ref{energy}) is also nonzero and energy 
consumption is also incorporated in the target minimization problem.
In this case, the value the minimization in Eq. (\ref{totalcost}) reached 
is the minimized  total monetary cost and energy consumption.
The third part is the penalty defined in Eq. (\ref{penalty}). The remaining 
data after deadline determines the penalty, which can not be directly 
controlled by MU. MU can only try to finish the data transmission
before the deadline to eliminate the penalty.
\fi 
\section{DQN Based Offloading Algorithm}\label{learningalgorithm}
 \indent In reinforcement learning, an agent makes optimal decision
 by acquiring knowledge of the unknown environment through
 reinforcement.  
 \ifmark
  \textcolor{red}{In our model in this paper, MU is the agent. 
  The state is the location and remaining data size. The action is to 
  choose cellular network, wireless LAN, or none of the two. 
  The negative reward is the monetary cost and energy consumption 
  at each time epoch. The MU's goal is to minimize the total 
  monetary cost and energy consumption over all the time epochs.}
  
  \textcolor{red}{One important difference that arises in 
  reinforcement learning and not in other kinds of learning is 
  the trade-off between \textit{exploration} 
  and \textit{exploitation}.  To minimize the total monetary cost and 
  energy consumption, MU must prefer actions that it has tried in the 
  past and found to be effective in reducing monetary cost and energy 
  consumption. But to discover such actions, it has to try actions that 
  it has not selected before. The MU has to \textit{exploit} what it already 
  knows in order to reduce monetary cost and energy consumption, 
  but it also has to \textit{explore} in order to make better action 
  selections in the future. }
  
  \textcolor{red}{Initially, the MU has no experience. The MU has to \textit{explore} 
  unknown actions to get experience of monetary cost and energy 
  consumption for some states. Once it gets the experiences, 
  it can \textit{exploit} what it already knows for the states but 
  keep \textit{exploring} at the same time. We use the parameter  
  $\epsilon$ in \textbf{Algorithm 1} to set the trade-off between 
  \textit{exploration} and \textit{exploitation}.}
 \else
  In our model in this paper, MU is the agent. 
  The state is the location and remaining data size. The action is to 
  choose cellular network, wireless LAN, or none of the two. 
  The negative reward is the monetary cost and energy consumption 
  at each time epoch. The MU's goal is to minimize the total 
  monetary cost and energy consumption over all the time epochs.
  
  One important difference that arises in 
  reinforcement learning and not in other kinds of learning is 
  the trade-off between \textit{exploration} 
  and \textit{exploitation}.  To minimize the total monetary cost and 
  energy consumption, MU must prefer actions that it has tried in the 
  past and found to be effective in reducing monetary cost and energy 
  consumption. But to discover such actions, it has to try actions that 
  it has not selected before. The MU has to \textit{exploit} what it already 
  knows in order to reduce monetary cost and energy consumption, 
  but it also has to \textit{explore} in order to make better action 
  selections in the future.
  
  Initially, the MU has no experience. The MU has to \textit{explore} 
  unknown actions to get experience of monetary cost and energy 
  consumption for some states. Once it gets the experiences, 
  it can \textit{exploit} what it already knows for the states but 
  keep \textit{exploring} at the same time. We use the parameter  
  $\epsilon$ in \textbf{Algorithm 1} to set the trade-off between 
  \textit{exploration} and \textit{exploitation}.
 \fi
 
 There are many methods for MU to get optimal policy
 in reinforcement learning. Our previous work \cite{ChengAPNOMS2016}
 adopted a Q-learning based approach, which is a kind of temporal
 difference (TD) learning algorithm \cite{RLBook}. TD learning algorithms
 require no model for the environment and are fully incremental.
 The core idea of Q-learning is to learn an action-value function
 that ultimately gives the expected MU's cost of taking a given
 action in a given state and following the optimal policy thereafter.
 The optimal action-value function follows the following
 \textit{optimality equation} (or \textit{Bellman equation})
 in Eq. (\ref{bellmaneq}) \cite{DPBook}.
 \begin{equation}\label{bellmaneq}
 \mathcal{Q}_t^*(\textbf{\textit{s}}_t,\textbf{\textit{a}}_{t})
  = \mathbb{E}_{\textbf{\textit{s}}_{t+1}}\big[ r_t(\textbf{\textit{s}}_t,\textbf{\textit{a}}_{t})
  + \gamma \min_{\textbf{\textit{a}}_{t+1}}\mathcal{Q}_t(\textbf{\textit{s}}_{t+1},\textbf{\textit{a}}_{t+1}) | \textbf{\textit{s}}_{t},\textbf{\textit{a}}_{t}\big]
\end{equation}
 where $\gamma$  is the discount factor in (0,1).
 The value of the action-value function is called Q-value.
 In Q-learning algorithm, the optimal policy can be easily
 obtained from optimal Q-value, $\mathcal{Q}_t^*(\textbf{\textit{s}},a)$,
 which is shown in Eq. (\ref{optimaldecision})
 \begin{equation}\label{optimaldecision}
   \phi_t^* = \arg\min_{\textbf{\textit{a}}_{t}\in\mathcal{A}}\mathcal{Q}_t^*(\textbf{\textit{s}}_t,\textbf{\textit{a}}_{t})
 \end{equation}
 Transition probability (or MU's mobility pattern, please 
 note that when we mention "unknown transition probability"
 in this paper, it is also means "unknown mobility pattern") 
 that is necessary in \cite{MDPOffloading2017} is no longer needed
 in Q-learning based offloading algorithm. However, there are
 three problems in the reinforcement learning algorithm
 in \cite{ChengAPNOMS2016}.
  \begin{table}[t]
\renewcommand{\arraystretch}{1}
\begin{tabular}{p{0.1cm}p{7.8cm}}
\hline

\hline
&\textbf{Algorithm 1}: DQN Based Offloading Algorithm  \\
\hline

\hline
1:& Initialize replay memory $D$ to capacity $N$\\
2:& Initialize action-value function $\mathcal{Q}$ with random parameters $\theta$\\
3:& Initialize target action-value function $\bar{\mathcal{Q}}$ with  parameters $\theta^-$\\
4: &Set $t:=1$, $\textbf{\textit{b}}_1:=\mathcal{B}$,  and set $l_1$ randomly.\\
5: &Set $\textbf{\textit{s}}_1=(l_1,\textbf{\textit{b}}_1)$\\
6: &\textbf{while} $t\le T$ and $b > 0$:\\
7: &\htab $l_t$ is determined from GPS\\
8: &\htab $rnd$ $\leftarrow$ random number in [0,1]\\
9: &\htab \textbf{if} $rnd$ $<$ $\epsilon$ :\\
10: &\htab\htab Choose action $a$ randomly\\
11: &\htab \textbf{else}: \\
12: &\htab\htab Choose action $a$ based on Eq. (\ref{optimaldecision2}) \\
13: &\htab \textbf{end if}\\
14: &\htab Set $\textbf{\textit{s}}_{t+1} = (l_t\textbf,  [\textbf{\textit{b}}_t-
\delta \textbf{\textit{a}}_{t,c}-\delta \textbf{\textit{a}}_{t,w}]^+)$\\ 
15: &\htab Calculate $r_t(\textbf{\textit{s}}_t,\textbf{\textit{a}}_{t})$ by Eq. (\ref{reward})\\
16: &\htab Store experience $(\textbf{\textit{s}}_t, \textbf{\textit{a}}_{t}, r_t,
\textbf{\textit{s}}_{t+1})$ in $D$\\
17: &\htab Sample random minibatch of $(\textbf{\textit{s}}_j, \textbf{\textit{a}}_{j}, r_j,
\textbf{\textit{s}}_{j+1})$ from $D$\\
18: &\htab \textbf{if} $j+1$ is termination:\\
19: &\htab \htab  Set $z_j$ = $r_j$\\
20: &\htab \textbf{else:} \\
21: &\htab \htab  Set $z_j$ = $r_j
  + \gamma \min_{\textbf{\textit{a}}_{j+1}}\mathcal{Q}_t(\textbf{\textit{s}}_{j+1},\textbf{\textit{a}}_{j+1}; \theta_j^{-})$\\
22: &\htab \textbf{end if} \\
23: &\htab Execute a gradient descent step on $(z_j-\mathcal{Q}_t(\textbf{\textit{s}}_{t},\textbf{\textit{a}}_{t}; \theta)^2$ with respect to parameters $\theta$.\\
24: &\htab Every C steps reset $\mathcal{\bar{Q}}$=$\mathcal{Q}$\\
25: &\htab Set $t:=t+1$\\
26: &\textbf{end while} \\
\hline

\hline
\end{tabular} \\
\end{table}
 \begin{itemize}
 \item (i) The state discretization induces error.  The state of
 remaining data is continuous, which is discretized in algorithms
 in \cite{ChengAPNOMS2016} and \cite{MDPOffloading2017}
 as well. One way to reduce the error is to use small granularity
 to discretize the remaining data, which increases the number
 of state.
 \item (ii) The large number of state makes it difficult to
 implement the Q-learning algorithm. The simplest way of
 implementation is to use  two-dimension table to store
 Q-value data, in which one of the dimensions indicates
 states and the other indicates actions. The method quickly
 becomes inviable with increasing sizes of state/action.
 \item (iii) The convergence rate of the algorithm is rather low.
 The algorithm begins to converge if MU experience many
 states and the agent does not have the ability to generalize its
 experience to unknown states.
 \end{itemize}
 Therefore, we propose to use DQN \cite{DeepRL} based
 algorithm to solve the problem in Eq.(\ref{totalcost}).
 In DQN based algorithm, DNN \cite{Deeplearning}
 is used to generalize MU's experience to predict the Q-value
 of unknown states.  Furthermore, the continuous state of
 remaining data is fed into DNN directly without discretization
 error.

 In DQN, the action-value function is estimated by a
 function approximator $\mathcal{Q}_t(\textbf{\textit{s}},a;\mathbf{\theta})$
 with parameters $\mathbf{\theta}$.
 Then MU's optimal policy is obtained from the following
 Eq.(\ref{optimaldecision2}) instead of Eq.(\ref{optimaldecision})
 \begin{equation}\label{optimaldecision2}
   \phi_t^* = \arg\min_{\textbf{\textit{a}}_{t}\in\mathcal{A}}\mathcal{Q}_t^*(\textbf{\textit{s}}_t,\textbf{\textit{a}}_{t}; \theta)
 \end{equation}
 A neural network function approximator with weights $\mathbf{\theta}$
 is called as a Q-network.  The Q-network can be trained by changing
 the parameters $\theta_i$ at iteration $i$ to decrease the mean-squared
 error in the Bellman equation, where the optimal
 target values in Eq.(\ref{bellmaneq}),
 $r_t(\textbf{\textit{s}}_t,\textbf{\textit{a}}_{t})
  + \gamma \min_{\textbf{\textit{a}}_{t+1}}\mathcal{Q}_t(\textbf{\textit{s}}_{t+1},\textbf{\textit{a}}_{t+1})$,
  are replaced by the approximate target values
  \begin{equation}
  z = r_t(\textbf{\textit{s}}_t,\textbf{\textit{a}}_{t})
  + \gamma \min_{\textbf{\textit{a}}_{t+1}}\mathcal{Q}_t(\textbf{\textit{s}}_{t+1},\textbf{\textit{a}}_{t+1}; \theta_i^{-})
  \end{equation}
  where $\theta_i^{-}$ is the parameter in the past iteration.

  The mean-squared error (or loss function) is defined as in Eq.(\ref{lossfunc}).
  \begin{equation}\label{lossfunc}
  L_i (\theta_i) = \mathbb{E}_{\textbf{\textit{s}}_t,\textbf{\textit{a}}_{t}, r_t, \textbf{\textit{s}}_{t+1}}\big[(z - \mathcal{Q}_t(\textbf{\textit{s}}_t,\textbf{\textit{a}}_{t};\mathbf{\theta}_i))^2\big]
  \end{equation}
  The gradient of loss function can be obtained by differentiation as follows.
  \begin{equation}\label{LFGradient}
  \begin{split}
  &\nabla_{\theta_i}  L_i (\theta_i)= \\
  &\mathbb{E}_{\textbf{\textit{s}}_t,\textbf{\textit{a}}_{t}, r_t, \textbf{\textit{s}}_{t+1}}\big[\big(z - \mathcal{Q}_t(\textbf{\textit{s}}_t,\textbf{\textit{a}}_{t};\mathbf{\theta}_i)\big) \nabla_{\theta_i}\mathcal{Q}_t(\textbf{\textit{s}}_t,\textbf{\textit{a}}_{t};\mathbf{\theta}_i)\big]
  \end{split}
  \end{equation}
  The gradient  $\nabla_{\theta_i}\mathcal{Q}_t(\textbf{\textit{s}}_t,\textbf{\textit{a}}_{t};\mathbf{\theta}_i)$ gives the direction to
  minimize the loss function in  Eq.(\ref{lossfunc}). The parameter
  are updated by the following rule in Eq.(\ref{updaterule})
  \begin{equation}\label{updaterule}
  \theta_{i+1} = \theta_{i+1} + \alpha \nabla_{\theta_i}  L_i (\theta_i)
  \end{equation}
  where $\alpha$ is the learning rate in (0,1).
  The proposed DQN based offloading algorithm is shown in Algorithm 1.
  As shown from line 16 to line 23 in Algorithm 1, MU's experience 
  $(\textbf{\textit{s}}_t, \textbf{\textit{a}}_{t}, r_t, \textbf{\textit{s}}_{t+1})$ 
  are stored in replay memory. Therefore, transition probability is no 
  longer needed. The Q-value is estimated from continuous state in
  line 21 without discretization. Therefore, there is no discretization 
  error.
  
  \ifmark
  \textcolor{red}{The mobile terminal has all the information needed 
  in reinforcement learning: the state, action, monetary cost and 
  energy consumption at each time $t$, and the goal. Specifically, 
  the mobile terminal has the location information from GPS, and 
  the keep recording the remaining data for each flow, then mobile 
  terminal has the information of state. The mobile terminal can 
  also detect the candidate actions--cellular network, wireless 
  LAN--at each location at each time epoch. Since the price of 
  cellular network and energy consumption for different data rate 
  is already known by MU, then the monetary cost  and energy 
  consumption information are also known to MU. The server just 
  provides data source for mobile terminal to download.}
  \else
The mobile terminal has all the information needed 
  in reinforcement learning: the state, action, monetary cost and 
  energy consumption at each time $t$, and the goal. Specifically, 
  the mobile terminal has the location information from GPS, and 
  the keep recording the remaining data for each flow, then mobile 
  terminal has the information of state. The mobile terminal can 
  also detect the candidate actions--cellular network, wireless 
  LAN--at each location at each time epoch. Since the price of 
  cellular network and energy consumption for different data rate 
  is already known by MU, then the monetary cost  and energy 
  consumption information are also known to MU. The server just 
  provides data source for mobile terminal to download.
  \fi

\section{Performance Evaluation}\label{performanceevaluation}
In this section, the performances of our \textit{proposed DQN} 
based offloading algorithm are evaluated by comparing them 
with dynamic programming based offloading algorithm (\textit{DP}) 
and heuristic offloading (\textit{Heuristic}) algorithm 
in our previous work \cite{MDPOffloading2017}.
\ifmark
\textcolor{red}{We employed the DP and Heuristic methods as 
comparative methods to show that our proposed reinforcement learning 
based algorithm is effective even if the MU's mobility pattern 
(or transition probability) from one place to another is unknown. }

\textcolor{red}{While the DP algorithm was proposed to get the 
optimal solution of the formulated Markov decision process (MDP) 
problem, the Heuristic algorithm was proposed to get near-optimal
solution of the same problem with low time-complexity. These 
two algorithms were based on the assumption that MU's transition 
probability from one place to another is known in advance.  
We try to show by comparison that our proposed DQN algorithm 
in this paper is valid even if the MU's transition probability is unknown.}

\textcolor{red}{Since the transition probability is necessary for 
DP and heuristic algorithms to work, we input some "incorrect" 
transition probability by adding some noise to the MU's "true"  
transition probability and check the performance.}\\
\else
{We employed the DP and Heuristic methods as 
comparative methods to show that our proposed reinforcement learning 
based algorithm is effective even if the MU's mobility pattern 
(or transition probability) from one place to another is unknown. }

{While the DP algorithm was proposed to get the 
optimal solution of the formulated Markov decision process (MDP) 
problem, the Heuristic algorithm was proposed to get near-optimal
solution of the same problem with low time-complexity. These 
two algorithms were based on the assumption that MU's transition 
probability from one place to another is known in advance.  
We try to show by comparison that our proposed DQN algorithm 
in this paper is valid even if the MU's transition probability is unknown.}

{Since the transition probability is necessary for 
DP and heuristic algorithms to work, we input some "incorrect" 
transition probability by adding some noise to the MU's "true"  
transition probability and check the performance.}\\
\fi
\indent We developed a simulator by Python 2.7, which can be downloaded
from URL link (https://github.com/aqian2006/\\
OffloadingDQN).\\
\begin{table}[th]
\caption{Energy vs. Throughput.}
\centering
\label{tableenergy}
\renewcommand{\arraystretch}{1}
\begin{tabular}{|c|c|}
\hline

\hline
Throughput (Mbps) & Energy (joule/Mb)\\
\hline
11.257 &  0.7107\\
\hline
16.529 &  0.484\\
\hline
21.433 &  0.3733\\
\hline

\hline
\end{tabular}
\end{table}
\indent A four by four grid is used in simulation.
Therefore, $L$ is 16. Wireless LAN APs are randomly
deployed in $L$ locations. The cellular usage price is
assumed as  1.5 yen/Mbyte. 
\ifmark
\textcolor{red}{It is assumed that each location is 
a square of 10 meters by 10 meters. The MU is an 
pedestrian and the time slot length is assumed 
as 10 seconds. The epsilon greedy parameter
$\epsilon$ is set to 0.08. }
\else
{It is assumed that each location is 
a square of 10 meters by 10 meters. The MU is an 
pedestrian and the time slot length is assumed 
as 10 seconds. The epsilon greedy parameter
$\epsilon$ is set to 0.08. }
\fi
$\textrm{Pr}(l|l)=0.6$ means that the probability that
MU stays in the same place from time $t$ to $t'$ is 0.6.
And MU moves to the neighbour location with equal 
probability, which can be calculated as
$\textrm{Pr}(l_{t+1}|l_t)=(1-0.6)/(\textrm{number of neighbors})$.
Because DP algorithm in \cite{MDPOffloading2017}
can not work without transition probability information,
we utilize the aforementioned transition probability but 
add some noise to the transition probability.
\ifmark
\textcolor{red}{
Please note that our proposed DQN based algorithm do 
not need the transition probability $\textrm{Pr}$.
The Pr is externally given to formulate the simulation 
environment. The less the Pr is, the more dynamic of MU is.  
High Pr value means the the  probability that MU stays a the 
same place is high.  If Pr is changed to small value,
the proposed DQN based algorithm is expected to have 
better performance. The reason is that in the DQN based 
algorithm, the exploration and exploitation is incorporated 
into the algorithm, which can adapt to environment changes.}
\else
Please note that our proposed DQN based algorithm do 
not need the transition probability $\textrm{Pr}$.
The Pr is externally given to formulate the simulation 
environment. The less the Pr is, the more dynamic of MU is.  
High Pr value means the the  probability that MU stays a the 
same place is high.  If Pr is changed to small value,
the proposed DQN based algorithm is expected to have 
better performance. The reason is that in the DQN based 
algorithm, the exploration and exploitation is incorporated 
into the algorithm, which can adapt to environment changes.
\fi
The average Wireless LAN throughput
$\gamma_{t,w}^l$ is assumed as 15 Mbps\footnote{We 
tested repeatedly with an iPhone 5s on the public wireless 
LAN APs of one of the biggest Japanese wireless carriers.
The average throughput was 15 Mbps. }, while
average cellular network throughput $\gamma_{t,c}^l$ is 
10 Mbps\footnote{We also tested with an iPhone 5s on 
one of the biggest Japanese wireless carriers' cellular 
network. We use the value 10 Mbps for average cellular 
throughput.}. We generate wireless LAN throughput for 
each AP from a truncated normal distribution, and the 
mean and standard deviation are assumed as 15Mbps 
and 6Mbps respectively.  The wireless LAN throughput
is in the range [9Mbps, 21Mbps]. Similarly, we generate 
cellular throughput from a truncated normal distribution, 
and the mean and standard deviation are assumed as 
10Mbps and 5Mbps respectively.  The cellular network
throughput is in range [5Mbps, 15Mbps].
$\sigma$ in Algorithm 1 is assumed as 1 Mbytes.
Each epoch lasts for 1 seconds. The penalty function is assumed as
$g(\textbf{\textit{b}}_t)$=$2\sum_{j\in \mathcal{M}} b_t^j$.
Please refer to Table \ref{tableParm} for the parameters used in
the simulation.\\
\begin{figure}[h]
   \centering
   \includegraphics[width=63mm]{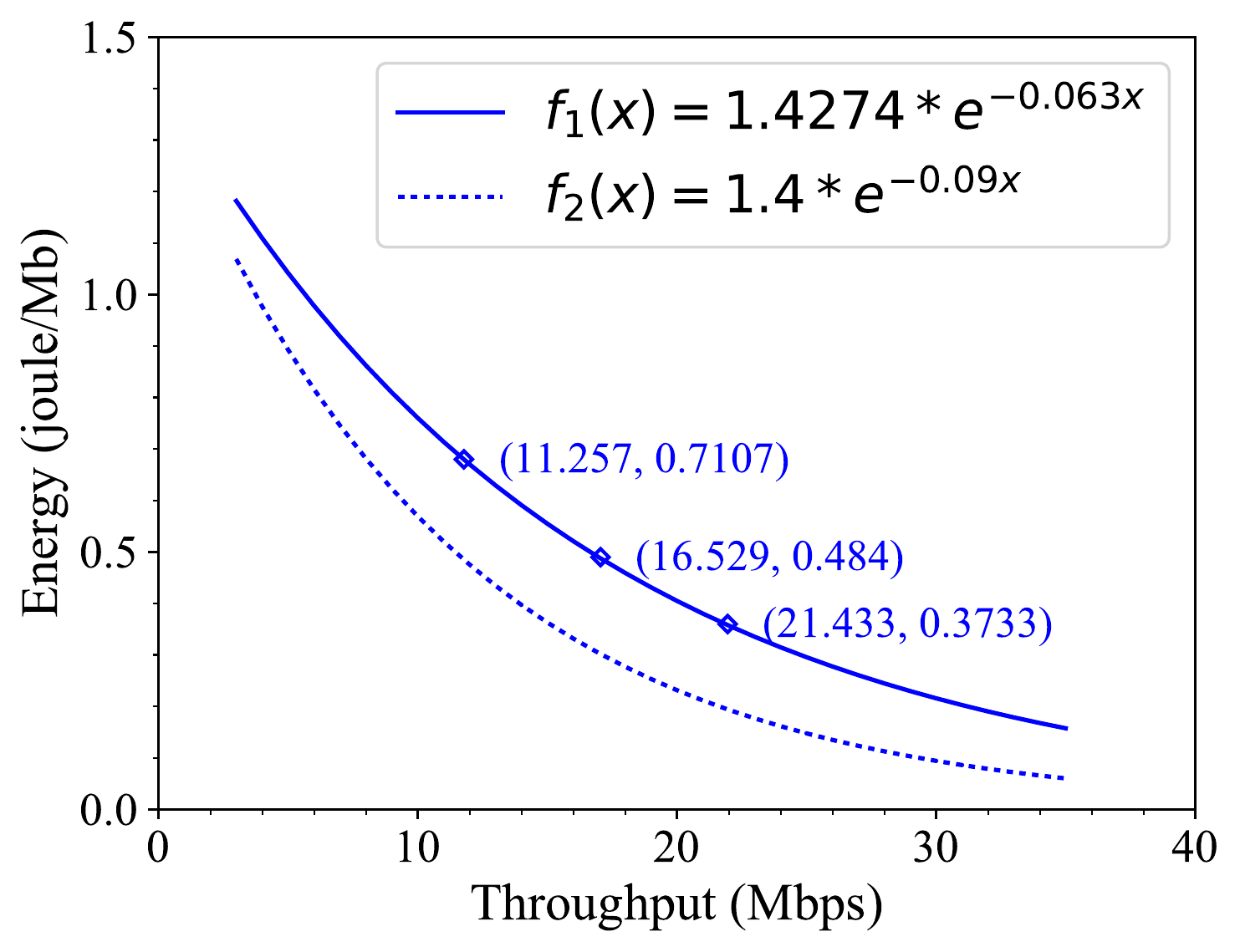}
   \caption{Energy consumption (joule/Mb) vs. Throughput (Mbps).}\label{energythroughput}
\end{figure}
\indent Because the energy consumption rate is a decreasing function of
throughput, we have the sample data from \cite{EnergyThroughput}
(see Table {\ref{tableenergy}). We then fit the sample data by
an exponential function  $f_1(x)=1.4274*\mathrm{e}^{-0.063x}$
as shown in Fig.  \ref{energythroughput}.
We also made a new energy-throughput
function as $f_2(x)=1.4*\mathrm{e}^{-0.09x}$, 
which is just lower than $f_1(x)$.
If we do not explicitly point that we use $f_2(x)$, 
we basically use $f_1(x)$.
Please note that the energy consumption rate of cellular 
and wireless LAN may be different for the same throughput, 
but we assume they are the same and use the same 
fitting function as in Fig. \ref{energythroughput}.
\iftable
\begin{table}[t]
\caption{Parameters in the simulation.}\label{tableParm}
\centering
\label{tableparameters}
\renewcommand{\arraystretch}{1}
\begin{tabular}{|c|c|}
\hline

\hline
Parameters & $\hspace{1cm}$Value$\hspace{1cm}$\\
\hline

\hline
 $L$ &16\\
\hline
\ifmark
 $\textbf{\textit{B}}$ &$\textbf{\textit{B}}=(400,600,800,1000)$ \textcolor{blue}
{Mbytes}\\
\else
 $\textbf{\textit{B}}$ &$\textbf{\textit{B}}=(400,600,800,1000)$ Mbytes\\
\fi
\hline
 $\mathcal{T}$ &$\mathcal{T}=(400,800,1200,1600)$\\
\hline
Number of wireless LAN APs &8\\
\hline
\ifmark
$\sigma$ & 1 \textcolor{blue}{Mbytes}\\
\else
$\sigma$ & 1 Mbytes\\
\fi
\hline
time slot & 10 seconds\\
\hline
$\epsilon$ & 0.08\\
\hline
average of $\gamma_{c}^l$ & 10 Mbps\\
\hline
standard deviation of $\gamma_{c}^l$ & 5 Mpbs\\
\hline
average of $\gamma_{w}^l$ & 15 Mbps\\
\hline
standard deviation of $\gamma_{w}^l$ & 6 Mpbs\\
\hline
$\textrm{Pr}({l|l})$ & 0.6\\
\hline
$\textrm{Pr}(l_{t+1}|l_t)$ & (1-0.6)/\#neigbour locations\\
\hline
$p_c$ & 1.5 yen per Mbyte\\
\hline
$g(\textbf{\textit{b}}_t)$ & $g(\textbf{\textit{b}}_t)$=$2\sum_{j\in \mathcal{M}} b_t^j$\\
\hline

\hline
\end{tabular}
\end{table}
\else
\fi

\begin{figure}[h]
   \centering
   \includegraphics[width=\figwidth]{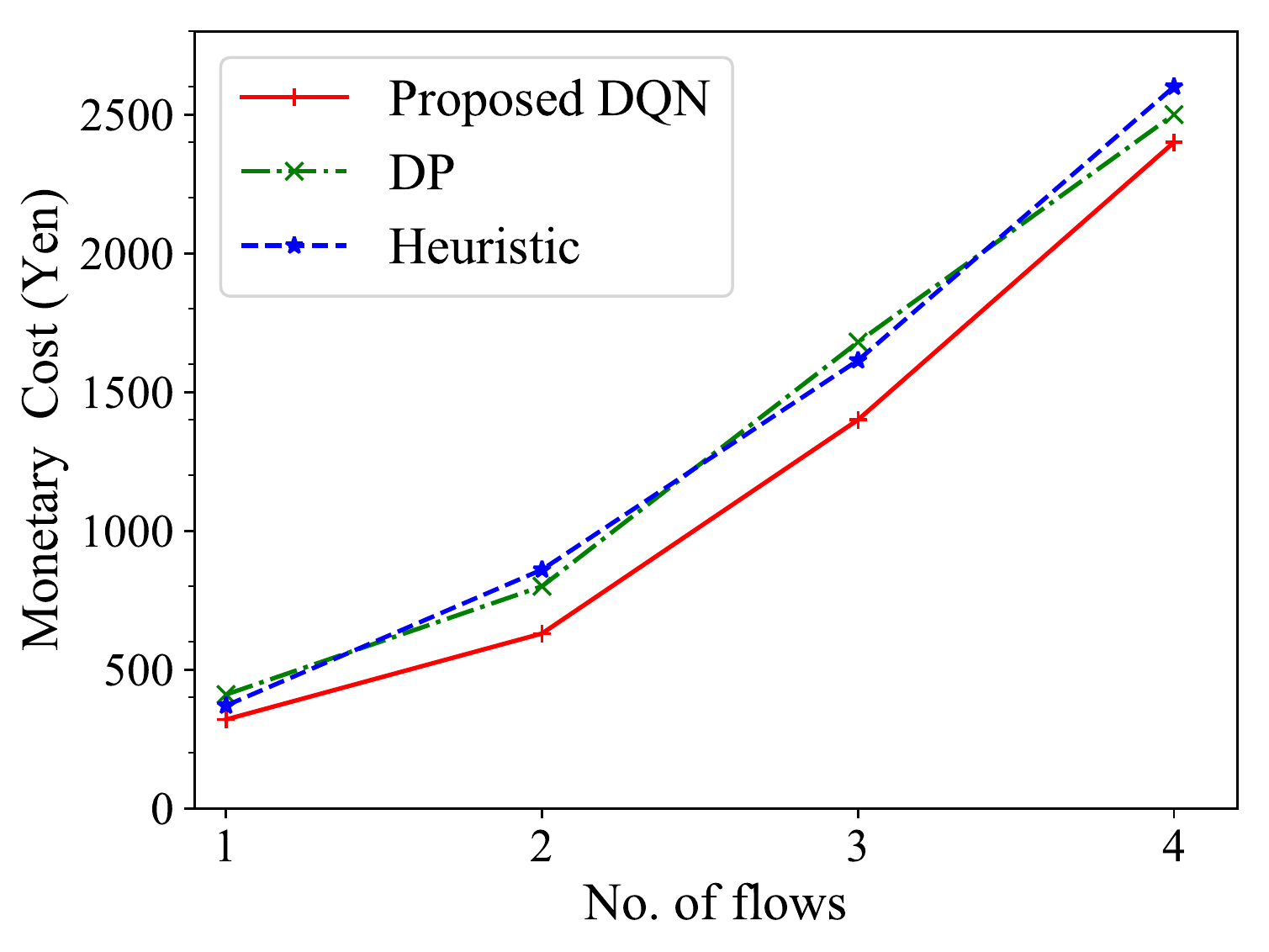}
   \caption{Monetary cost (yen) vs. No. of flows.}\label{FigCostFlows}
\end{figure}

\begin{figure}[h]
   \centering
   \includegraphics[width=\figwidth]{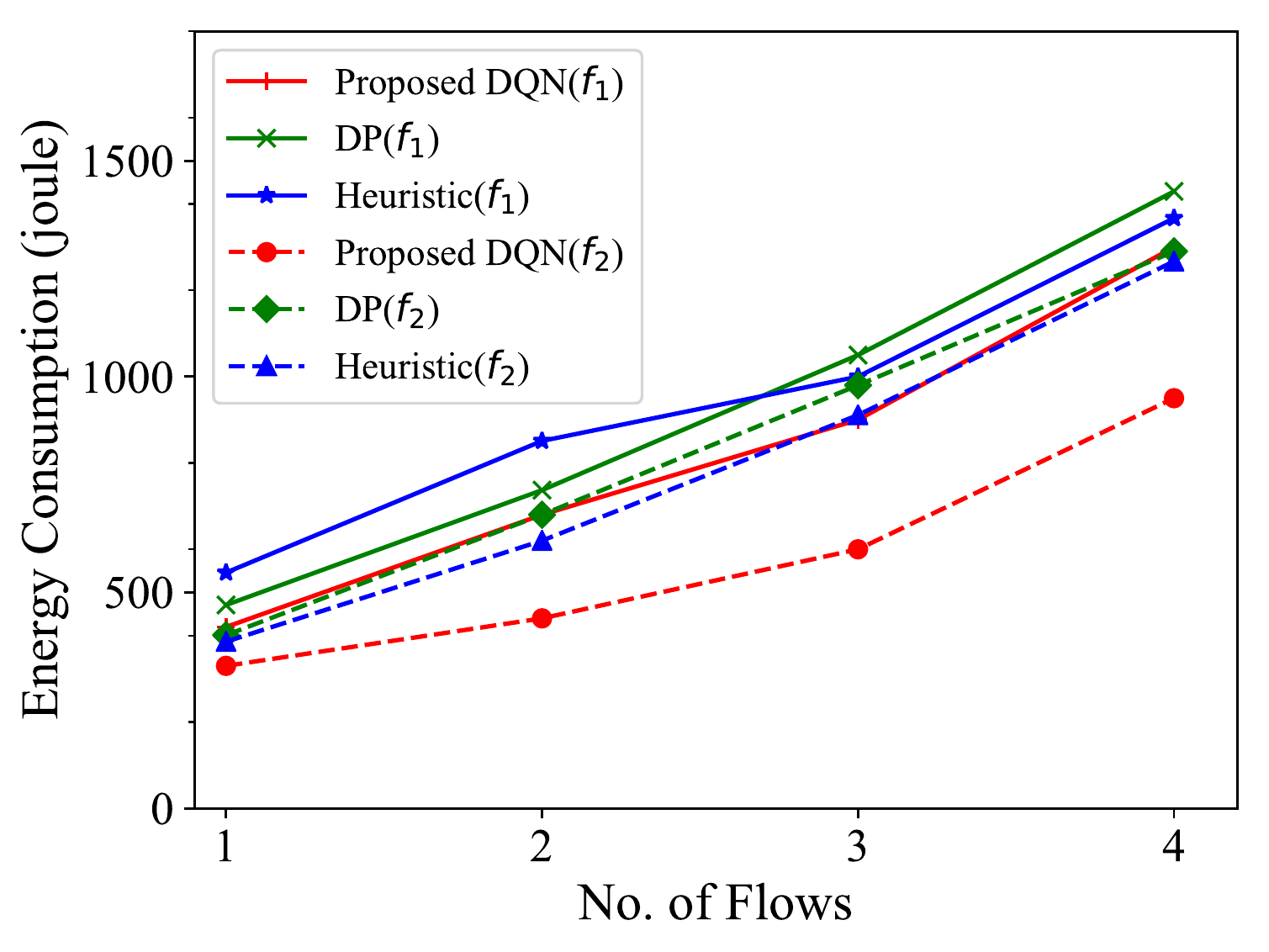}
\caption{Energy consumption (joule) vs. No. of flows with different energy-throughput functions $f_1$ and $f_2$.}\label{FigEnergyCompFLows}
\end{figure}

Fig.\ref{FigCostFlows} shows the comparison of monetary cost 
among \textit{Proposed DQN}, \textit{Heuristic} and \textit{DP}
algorithms with different number of flows.
The monetary cost of all three algorithms increases with the 
number of flows. 
\ifmark
\textcolor{red}
{Please note that we fixed the number of APs to 8 in Table 4.}
\else
{Please note that we fixed the number of APs to 8 in Table 4.}
\fi

The monetary cost of \textit{Proposed DQN} 
is lower than \textit{DP} and \textit{Heuristic}. And the \textit{Heuristic} 
sometimes performance better than \textit{DP}.
The reason is that the \textit{DP} algorithm with incorrect 
transition probability cannot obtain optimal policy,
while our \textit{Proposed DQN} can learn how to choose 
optimal policy with unknown transition probability.

Fig.\ref{FigEnergyCompFLows} shows the comparison of the 
energy consumption among \textit{Proposed DQN}, \textit{Heuristic}
and \textit{DP} algorithms with different number of flows.  
Two energy-throughput functions $f_1(x)$ and $f_2(x)$ are used. 
The overall energy consumption with $f_1(x)$ is greater than 
that with $f_2(x)$. The reason is that  energy consumption rate 
of $f_1(x)$ is much higher than that of $f_2(x)$ as shown in
Fig. \ref{energythroughput}. The performance of \textit{Proposed DQN} 
algorithm is the best with each energy-throughput function.
The reason is that the \textit{Proposed DQN} algorithm leans 
how to act optimally while both \textit{Heuristic} and \textit{DP} algorithms
can not act optimally without correct transition probability.
 \begin{figure}[h]
   \centering
   \includegraphics[width=\figwidth]{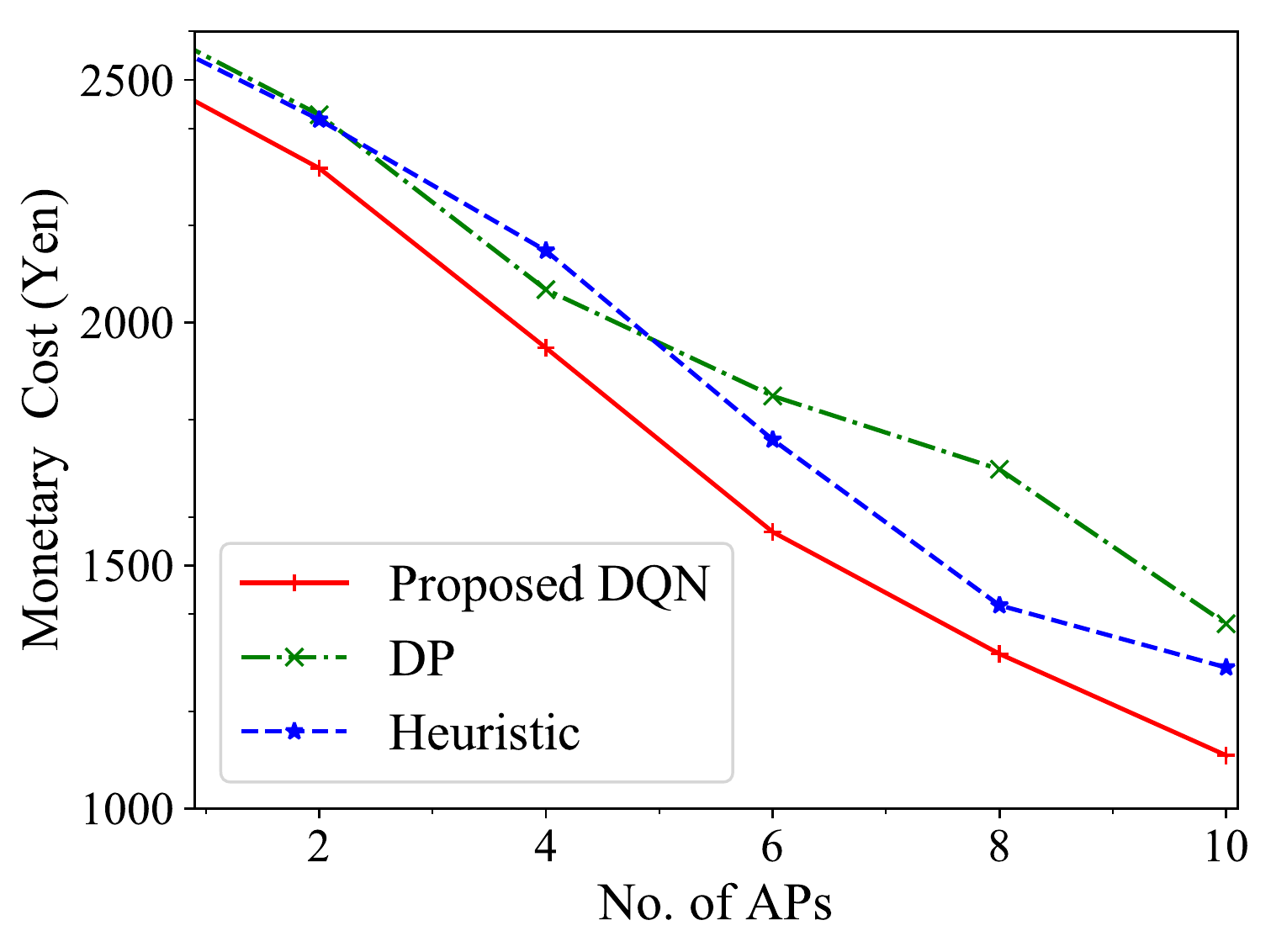}
   \caption{Monetary cost (yen) vs. No. of APs.}\label{FigCostAPs}
\end{figure}

\begin{figure}[h]
   \centering
   \includegraphics[width=\figwidth]{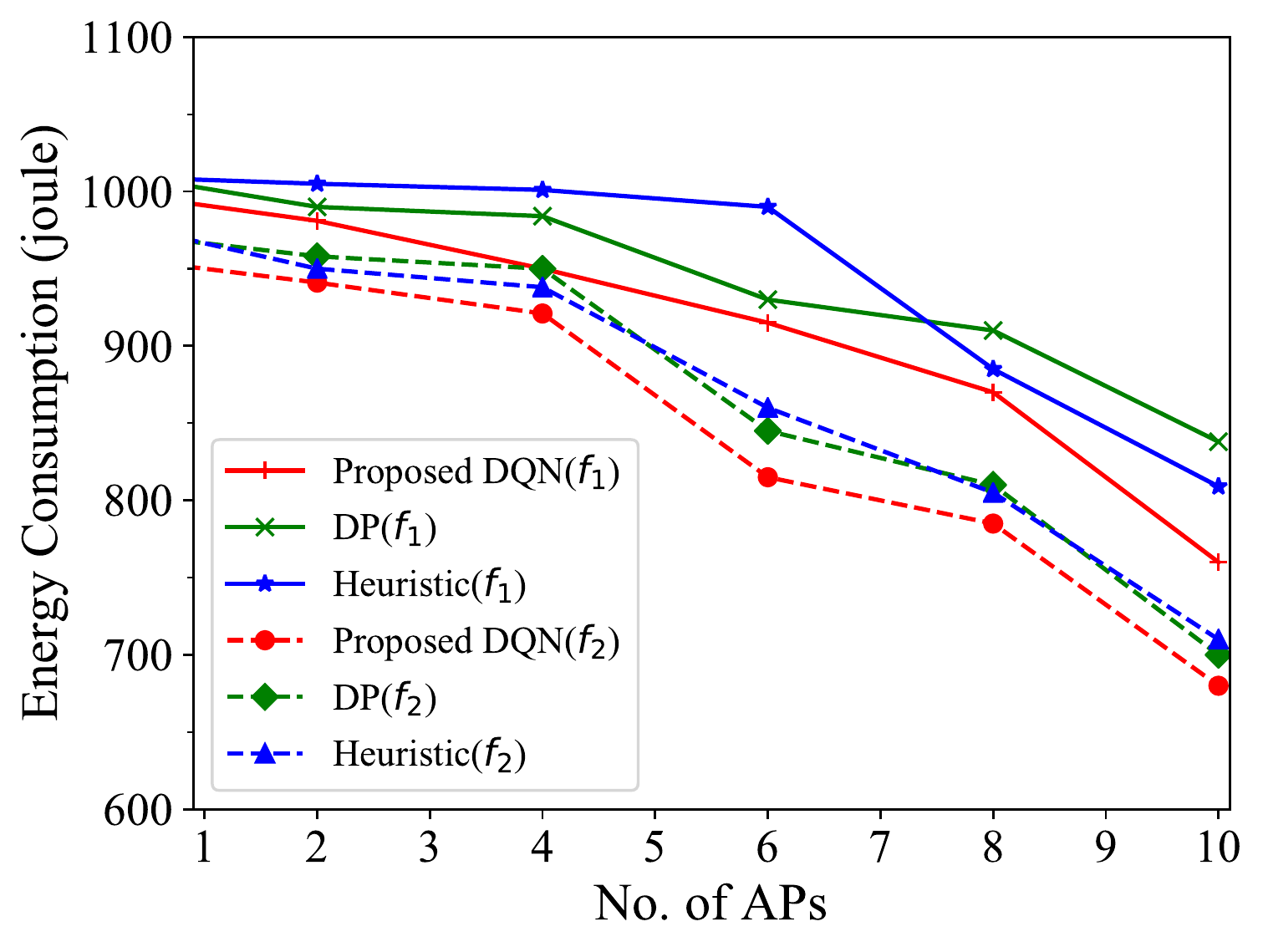}
\caption{Energy consumption (joule) vs. No. of APs with different 
energy-throughput functions $f_1$ and $f_2$.}\label{FigEnergyAPs}
\end{figure}

\begin{figure}[h]
   \centering
   \includegraphics[width=\figwidth]{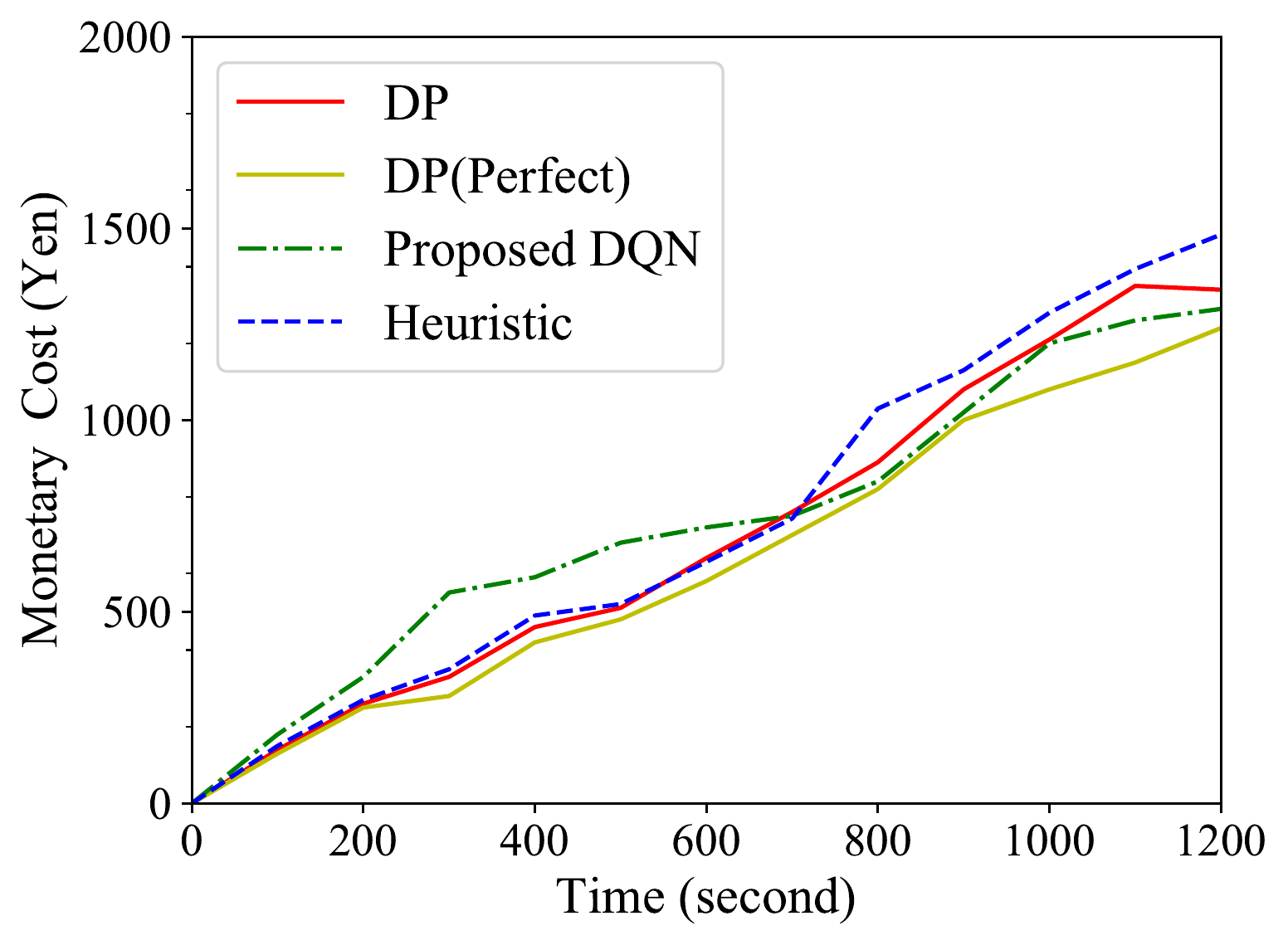}
   \caption{Monetary cost (yen) vs. Epochs.}\label{FigCostTime}
\end{figure}
\begin{figure}[h]
   \centering
   \includegraphics[width=\figwidth]{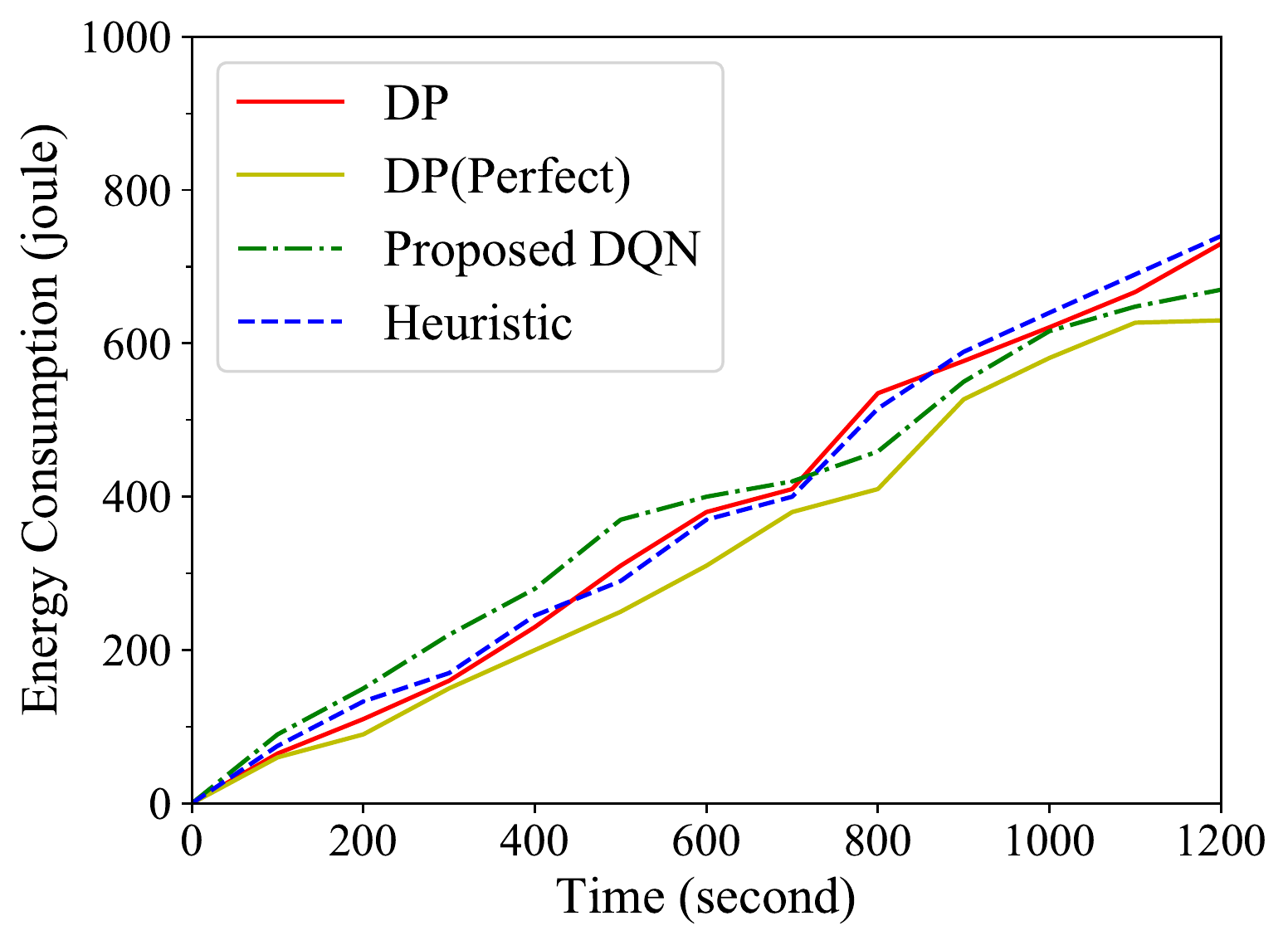}
\caption{Energy consumption (joule) vs. Epoch with $f_1$.}\label{FigEnergyTime}
\end{figure}

Fig.\ref{FigCostAPs} shows the comparison of monetary cost 
among \textit{Proposed DQN}, \textit{Heuristic} and \textit{DP} algorithms 
with different number of APs.
\ifmark
\textcolor{red}
{Please note that we fixed the number of flows to the first three
in Table 4.}
\else
{Please note that we fixed the number of flows to the first three
in Table 4.}
\fi
It can be seen that the monetary cost of \textit{Proposed DQN} 
algorithm is lowest. The reason is also that both both \textit{Heuristic} 
and \textit{DP} algorithms can not find the optimal policy with 
incorrect transition probability.
With a large number of wireless LAN APs deployed, the chance
of using cheap wireless LAN increases. Then, the MU can reduce 
his monetary cost by using cheap wireless LAN. Therefore, all three 
algorithms' monetary costs decreases with the number of APs.

Fig.\ref{FigEnergyAPs} shows how the MU's energy consumption 
changes with the number of deployed APs under the two 
energy-throughput functions $f_1(x)$ and $f_2(x)$.
Similar to Fig.\ref{FigEnergyCompFLows}, the performance of 
\textit{Proposed DQN} algorithm is the best with either $f_1(x)$ or $f_2(x)$.  
It shows that the energy consumptions of all three algorithms just  
slightly decrease with the number of APs. 
The reason is that the energy consumption depends
on the throughput. The larger the throughput, the lower is the 
energy consumption. With large number of wireless LAN APs, 
the MU has more chance to use wireless LAN with high throughput 
since the average throughput of a wireless LAN is assumed
as higher than that of cellular network (see Table \ref{tableParm}). \\
\ifmark
\textcolor{red}{
\indent Fig.\ref{FigCostTime} and Fig.\ref{FigEnergyTime} shows 
how monetary cost and energy consumption changes with 
time  among \textit{Proposed DQN}, \textit{Heuristic} and \textit{DP} 
algorithms with different number of APs.
It is obvious that it takes time for the \textit{Proposed DQN} to 
learn. The performance is not so good as \textit{Heuristic} and \textit{DP}.
But the performance of \textit{Proposed DQN} become better and 
better with time goes by.  We also have show the performance 
with perfect MDP transition probability of DP algorithm, it shows 
that the performance of \textit{DP} is best.}
\else
\indent Fig.\ref{FigCostTime} and Fig.\ref{FigEnergyTime} shows 
how monetary cost and energy consumption changes with 
time  among \textit{Proposed DQN}, \textit{Heuristic} and \textit{DP} 
algorithms with different number of APs.
It is obvious that it takes time for the \textit{Proposed DQN} to 
learn. The performance is not so good as \textit{Heuristic} and \textit{DP}.
But the performance of \textit{Proposed DQN} become better and 
better with time goes by.  We also have show the performance 
with perfect MDP transition probability of DP algorithm, it shows 
that the performance of \textit{DP} is best.}
\fi

\ifmemo
\begin{table}[th]
\textcolor{red}{\caption{Figures of simulation section.}}
\centering
\label{simulatioin-todo}
\renewcommand{\arraystretch}{1}
\begin{tabular}{|c|c|c|c|}

\hline

\hline
\textcolor{red}{No.}& x-axis & y-axis & parameter\\
\hline
\textcolor{red}{1}& energy preference ($\theta$)& energy consumption & \#flows, AP coverage\\
\hline
\textcolor{red}{2} & \#flow & monetary cost & $\theta$, AP coverage\\
\hline
\textcolor{red}{3} & \#flow & energy consumption & $\theta$, AP coverage\\
\hline
\textcolor{red}{4} & AP coverage & monetary cost & \#flow, $\theta$\\
\hline
\textcolor{red}{5} & AP coverage & energy consumption & \#flow, $\theta$\\
\hline
\textcolor{red}{6} & deadline & monetary cost & \#flow, $\theta$, AP coverage\\
\hline
\textcolor{red}{7} & deadline & energy consumption & \#flow, $\theta$, AP coverage\\
\hline
\textcolor{red}{8} & deadline & prob. completion & \#flow, $\theta$, AP coverage\\

\hline

\hline

\end{tabular}
\end{table}
\else
\fi

\section{Conclusion}\label{conclusion}
In this paper, we studied the multi-flow mobile data offloading problem
in which a MU has multiple applications that want to download data
simultaneously with different deadlines. We proposed a DQN
based offloading algorithm to solve the wireless LAN offloading 
problem to minimize the MU's monetary and energy cost. 
The proposed algorithm is effective even if the MU's mobility pattern 
is unknown.  The simulation results have validated our proposed 
offloading algorithm.

This work assumes that the MNO adopts usage-based pricing, in
which the MU paid for the MNO in proportion to data usage.
In the future, we will evaluate other usage-based pricing variants
like tiered data plan, in which the payment of the MU is a step
function of data usage. And we will also use time-dependent pricing
(TDP) we proposed in  \cite{TDPDuopolyNSP}\cite{TimeDependentPriceOligopolyNSP}, without changing
the framework and algorithms proposed in this paper.


\bibliographystyle{IEEEtran}
\bibliography{reference}


%





\ifCLASSOPTIONcaptionsoff
  \newpage
\fi

\end{document}